\documentclass[preprint,12pt, superscriptaddress]{revtex4}
\usepackage{graphics,epsfig}
\usepackage{epstopdf}
\usepackage{graphicx}
\usepackage{dcolumn}
\usepackage{amsmath}
\usepackage{epstopdf}
\begin{document}
\title{Parameter estimation of the Bardeen-Kerr black hole in cloud of strings using shadow analysis }

\author{Bijendra Kumar Vishvakarma}
\email{bkv1043@gmail.com}
\affiliation{Department of Physics, Institute of Science, Banaras Hindu University, Varanasi-221005, India}

\author{Dharm Veer Singh}
\email{veerdsingh@gmail.com}
\affiliation{Department of Physics,
Institute of Applied Science and Humanities, GLA University, Mathura-281406, 
India  }

\author{Sanjay Siwach}
\email{sksiwach@hotmail.com}
\affiliation{Department of Physics, Institute of Science, Banaras Hindu University, Varanasi-221005, India}
 
\begin{abstract}
We consider the rotating generalization of the Bardeen black hole solution in the presence of cloud of strings (CoS). The parameter space for which the black hole horizon exists is determined. We also study the static limit surface and the ergo-region in the presence of the CoS parameter. We consider photon orbits and obtain the deformation of black hole shadows due to rotation for various values of CoS parameter. The shadow deformation is used to determine the black hole spin for different values of the black hole parameters.  
\end{abstract} 

\maketitle

\section{\label{sec:level1}Introduction}
Black holes provide an interesting laboratory to test the predictions of the General Theory of Relativity as well as that of theories beyond General Relativity. The measurement of shadow size using the Event Horizon Telescope recently has opened up the possibility of determining the black hole parameters and future observations should be precise enough to make a distinction of black holes from different theories. A class of theories for which black hole solutions have been obtained in recent years are those with a non-linear electrodynamics source \cite{frad}. The recent interest in these solutions lies due to the absence of singularities for these solutions \cite{Ghosh:2018bxg,Singh:2022xgi, Singh:2017bwj,Maluf:2018lyu,bambi}. The theories of inflation and quantum gravity indicate the possibility of the existence of primordial black holes formed by self-gravitating magnetic mono-poles \cite{abg11,abg}. These black holes may have survived due to their topological stability and can provide clues about the observables in the early universe. They belong to Bardeen type space-time \cite{bard} and its generalizations \cite{dvs19,Singh:2020xju,k5,Singh:2019wpu,fr1,Singh:2017qur, Kumar:2018vsm,Singh:2020rnm,Yan:2023pxj,Singh:2022dth,Jafarzade:2020ova,Ghosh:2022gka, Belhaj:2020nqy}. 


Black holes are also investigated in non-trivial space-time e.g. cloud of strings (CoS) in order to mimic the early universe environment \cite{l1,l2,xu19}. In this context, the black hole solutions are investigated in CoS, and their shadows are constructed using numerical methods. These solutions are not asymptotically flat and provide new examples of space-time with this property. The generalization of these black holes to include the effects of Bardeen type non-linear electrodynamics (NED) was achieved recently \cite{Rodrigues:2022a,Rodrigues:2022b} (see also \cite{Belhaj:2022kek, Singh:2020nwo,Singh:2022ycn}). The solutions correspond to that of a self-gravitating magnetic mono-pole and may provide an opportunity to explore the black holes produced in the early universe. Their shadows and quasi-normal modes of this solution were also investigated recently \cite{bkv}.


Recently, the method of estimating the distortion of black hole shadow from circular shapes is proposed by Hioki and Meeda \cite{Hioki:2009na} and its generalizations in static \cite{EslamPanah:2020hoj, Sau:2022afl,Belhaj:2020okh,72,Kamruddin:2013iea} and rotating black holes are also considered \cite{Abdujabbarov:2016hnw,Ghosh:2020ece,Ahmed:2022qge,Kumar:2020owy,Jana:2023sil,Tan:2023ngk,Kumar:2020hgm}. The spatial angular resolution of VLBI radio observation is now below the horizon radius of super-massive black holes viz. Sgr A* and M87 \cite{Afrin:2021imp,Moffat:2019uxp,Kumar:2018ple,Pantig:2022qak,Psaltis:2018xkc,Banerjee:2022iok}. 
 This has opened up the possibility of determination of the parameter of the astrophysical black holes using black hole shadows  \cite{Afrin:2023uzo,KumarWalia:2022aop,Afrin:2022ztr,Afrin:2021wlj,Pulice:2023dqw}.

In this paper, we consider the rotating generalization of the Bardeen black hole in CoS. The rotating generalizations provide a unique opportunity to capture several observable features that are absent for charged black holes e.g. shape deformation of shadows. We calculate the range of parameters for which the horizon exists. The ergo-region and shadows are also plotted for different sets of parameters. The shadows around rotating black holes can be used to determine the parameters like spin and mass of the black holes. We use this to obtain the spin of the black for different values of the CoS parameter. The dependence of shadow radius on spin and CoS parameters is presented. The shape deformation parameters are obtained as a function of shadow radius.

The paper is organized as follows. In section II we review the Letelier-Bardeen black hole and present the rotating generalization using the Newman-Janus procedure. The horizon exists for a constrained set of black hole parameters only and we obtain this limit on parameter space numerically. The ergo-region is obtained in section III. In section IV, we consider the motion of mass-less particles (photons) around the black hole space-time and obtain the shadows for a permissible set of parameters. The distortion of black hole shadows from circular geometry is used to determine the black hole parameters in section VI. We summarise our results in the concluding section.   
 

\section{Letelier-Bardeen-Kerr black hole}

Let us consider the action of Einstein's gravity coupled with a  NED and cloud of strings source,  
\begin{equation}
S=\int d^4x \sqrt{-g}\left[R+{\cal L}_{NED}+{\cal L}_{cs}\right],
\label{action}
\end{equation}
where $R$ is the scalar curvature, ${\cal L}_{NED}$ and ${\cal L}_{CS}$ respectively are the Lagrangian density of the nonlinear source and CoS source. The equations of motion are obtained by varying the action with respect to metric tensor, $g_{\mu\nu}$ and electromagnetic potential, $A_{\mu}$, and can be written in the form,  
\begin{eqnarray}
&& R_{\mu\nu}-\frac{1}{2}g_{\mu\nu}R=T_{\mu\nu}^{NED}+T_{\mu\nu}^{CS},\\
&& \nabla_{\mu}\left(\frac{\partial {\cal L}}{\partial F}F^{\mu\nu}\right)=0
\label{eom}
\end{eqnarray}

where, we consider the Lagrangian density of the non-linear electrodynamics, which is taken as a function of $F=F_{ab}F^{ab}$ and specifically we consider the Bardeen type source given as,  \cite{Singh:2020xju,Singh:2017qur}

\begin{equation}
{{{\cal L}(F)}}= \frac{3}{2sg^2}\left(\frac{\sqrt{2g^2F}}{1+\sqrt{2g^2F}}\right)^{\frac{5}{2}}
\label{nonl1}
\end{equation}

 where $M$ and $g$ are the parameters to be identified with magnetic monopole charge and mass and $s=g/2M$.  The energy-momentum tensor can  be obtained from equation (\ref{nonl1}) and is given as,  

\begin{eqnarray}
&&  T_{ab}^{NED}=2\left[\frac{\partial {{L(F)}}}{\partial F}F_{a c}F_{\nu}^{c}-\tilde g_{ab}{{L(F)}}\right],
\label{emt}
\end{eqnarray}

The cloud of strings term in the action is given by the Nambu-Goto action and the energy-momentum tensor is given by, \cite{xu19} 

\begin{equation}
T^{\mu \nu}  = \frac{\rho \Sigma^{\mu \rho} \Sigma_{\rho}^{\phantom{\rho} \nu}}{\sqrt{-\gamma}}.
\end{equation}
where $\rho$ is the density and $\gamma$ is the induced metric on the worldsheet.  The $\Sigma^{\mu \nu}$ is a bivector given by,  
\begin{equation}
\label{eq:bivector}
\Sigma^{\mu \nu} = \epsilon^{a b} \frac{\partial x^{\mu}}{\partial \lambda^{a}} \frac{\partial x^{\nu}}{\partial \lambda^{b}},
\end{equation}
where $\epsilon^{a b}$ being the Levi-Civta tensor.

Let us consider the ansatz for the static spherically symmetric space-time, given by the line element
\begin{equation}
ds^2 = -f(r) dt^2+ \frac{1}{f(r)} dr^2 + r^2 d\Omega^2,
\label{met1}
\end{equation}
where $d\Omega^2=d\theta^2+\sin^2\theta d\phi^2$. We take the following form of the metric function,  
\begin{equation}
f(r)=1-\frac{2m(r)}{r}.
\label{met2}
\end{equation}

For magnetically charged black holes, the non-linear electrodynamics field strength can be taken in the form, of $F_{\theta\phi}=2g\sin\theta$ and the non-vanishing components of energy-momentum tensor (EMT) are given by,
\begin{eqnarray}
{T}^t_t = { T}^r_r =\frac{8M g^2}{(r^2+g^2)^{5/2}}+\frac{b}{r^2},
\label{emt1}
\end{eqnarray}
where ($M$) is black hole mass and ($a$) is the CoS parameter. The equations of motion give,
 \begin{eqnarray}
m'(r) =\frac{8M g^2}{(r^2+g^2)^{5/2}}+\frac{b}{r^2},
\label{emt1}
\end{eqnarray}
which can be integrated to give the black hole solution \cite{Rodrigues:2022a,Rodrigues:2022b}  

\begin{equation}
ds^2=-\left(1-\frac{2Mr^2}{(r^2+g^2)^{3/2}}-b\right) dt^2+\frac{1}{\left(1-\frac{2Mr^2}{(r^2+g^2)^{3/2}}-b\right)}+ r^2 d\theta^2+r^2\sin^2\theta d\phi^2
\label{solution}
\end{equation}
 
This is a Letelier-Bardeen-like black hole characterized by its mass ($M$), magnetic monopole charge ($g$), and a CoS parameter ($b$). To obtain the rotating counterpart of the black hole, we employ the Newman-Janus procedure and get the metric of the Letelier-Bardeen-Kerr black hole.
\begin{eqnarray}
 ds^{2}=-\left(1-\frac{b r^2+\frac{2M r^4}{(r^2+g^2)^{3/2}}}{\Sigma}\right)dt^{2}-\frac{2a\sin^{2}\theta}{\Sigma}\left(r^{2}+a^{2}-\Delta\right) dtd\phi+\frac{\Sigma}{\Delta}dr^{2}+\nonumber\\
\Sigma \,d\theta^{2}  +\frac{\sin^{2}\theta}{\Sigma}((r^{2}+a^{2})^{2}-\Delta ~ a^{2}\sin^{2}\theta)d\phi^{2},
\label{bhs}
\end{eqnarray}
where
\begin{eqnarray}
 \Delta=(1-b) r^{2}+a^{2}-\frac{2Mr^{4}}{(r^2+g^2)^{\frac{3}{2}}} \qquad\text{and}\qquad\Sigma = r^{2}+a^{2} \cos^{2}\theta
\end{eqnarray}
Eq. (\ref{bhs}) represents the rotating counterpart of Latelier- Bardeen black hole space-times in the Boyer-Lindquist coordinates. The spin parameter ($a=J/M$) is the ratio of the angular momentum ($J$) and ADM mass ($M$) of the rotating black hole. The solution (\ref{bhs}) goes over to the Bardeen-Kerr black hole in the absence of CoS parameters.

Next, we investigate the horizon structure of the black hole solution (\ref{bhs}) that corresponds to the space-time points $ (g^{rr}=\Delta=0)$:
\begin{equation}
  (1-b) r^{2}+a^{2}-\frac{2Mr^{4}}{(r^2+g^2)^{\frac{3}{2}}}|_{r_+}=0. 
  \label{hor}
\end{equation}

\begin{figure*}[ht]
\begin{tabular}{c c c c}
\includegraphics[width=.33\linewidth]{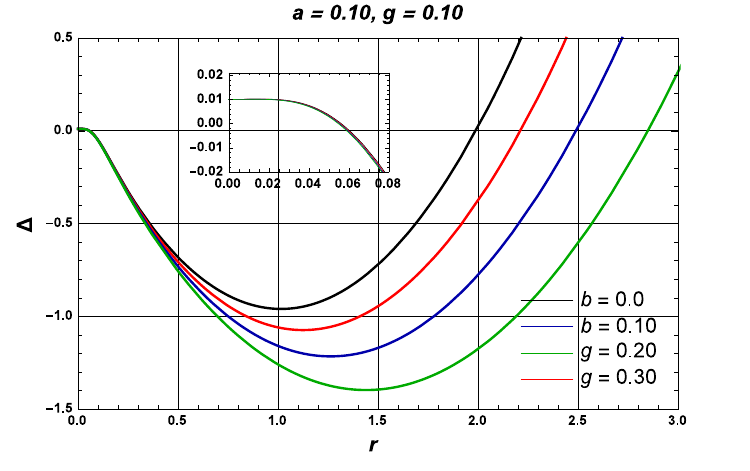}
\includegraphics[width=.33\linewidth]{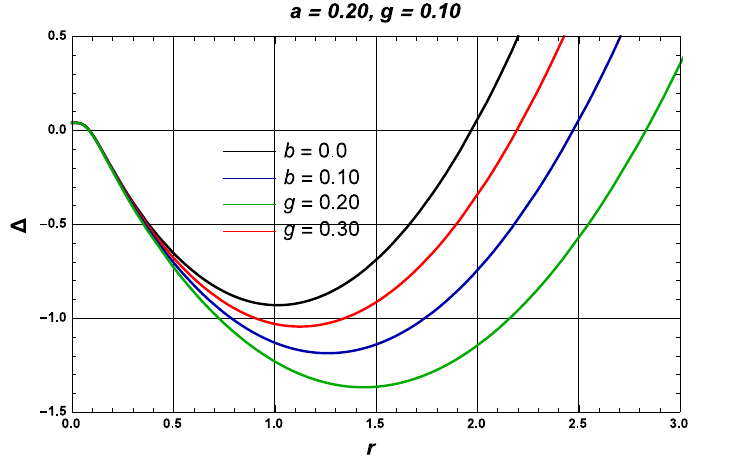}
\includegraphics[width=.33\linewidth]{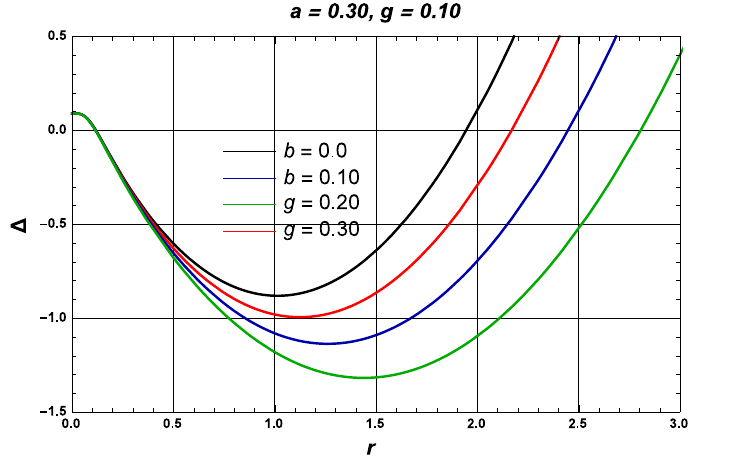}\\
\includegraphics[width=.33\linewidth]{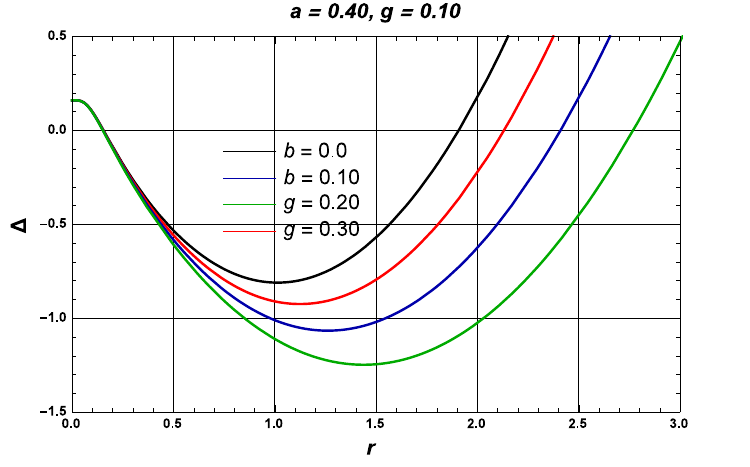}
\includegraphics[width=.33\linewidth]{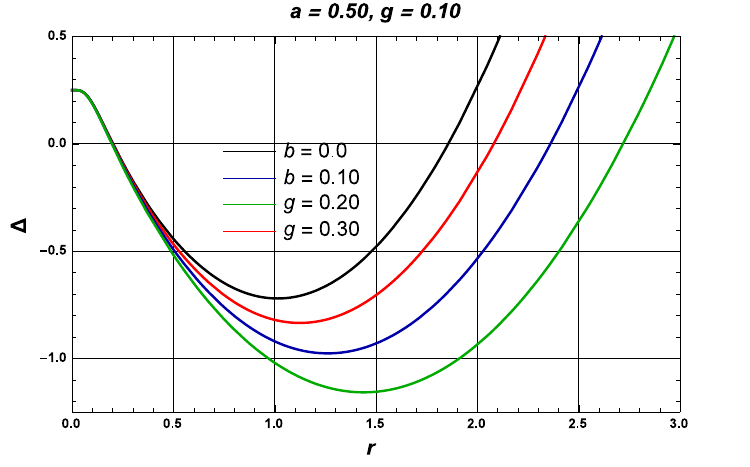}
\includegraphics[width=.33\linewidth]{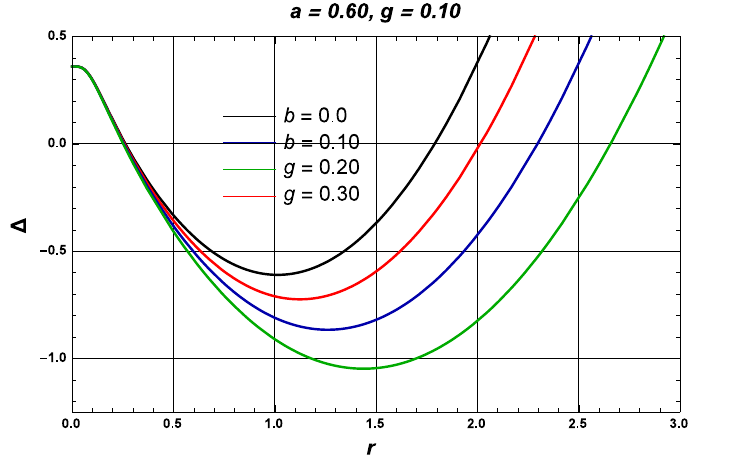}\\
\includegraphics[width=.33\linewidth]{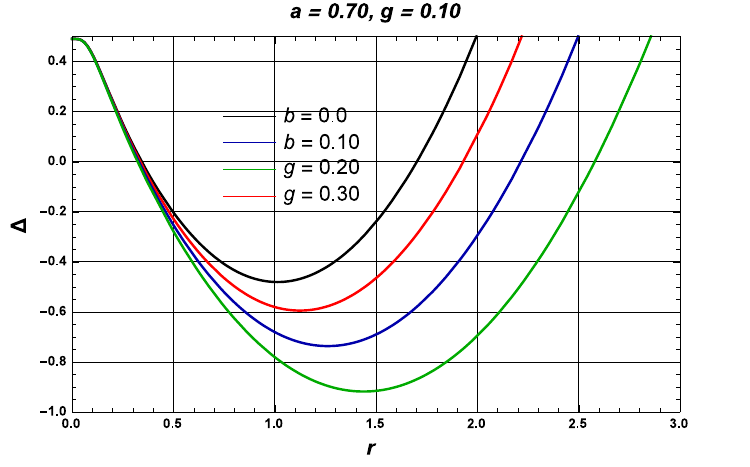}
\includegraphics[width=.33\linewidth]{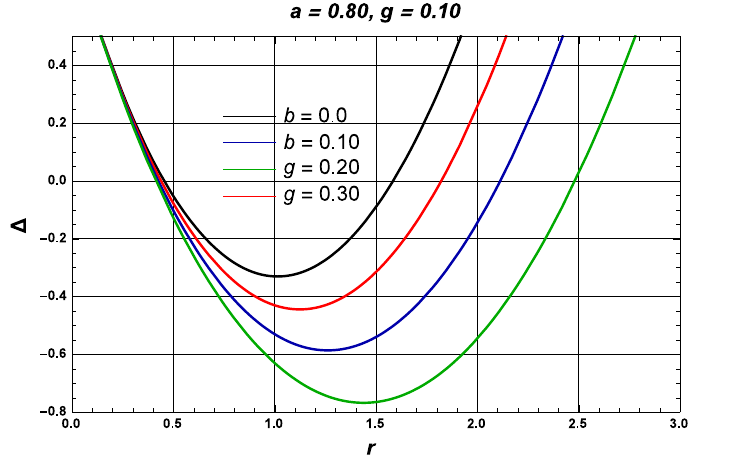}
\includegraphics[width=.33\linewidth]{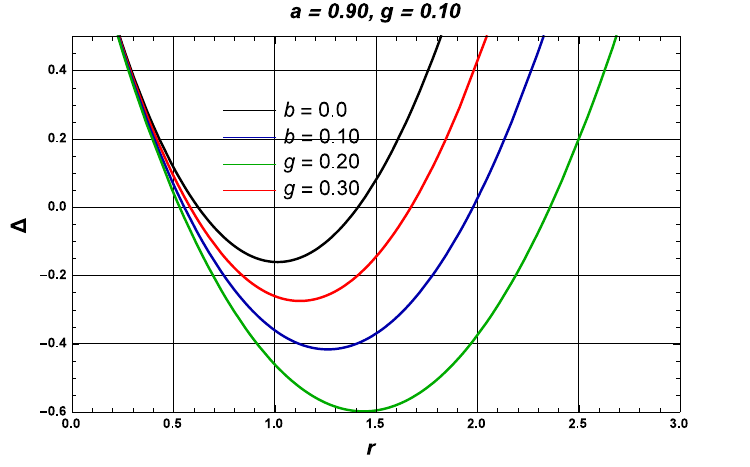}
\end{tabular}
\caption{Metric function ${\Delta}(r)$ vs  $r$ for different values of the spin parameter and CoS parameter with a fixed value of black hole mass ($M=1$) and magnetic monopole charge ($g=0.1$).}
\label{fig:ep}
\end{figure*}

 The Eq. (\ref{hor}) gives the location of the black hole horizons, which can not be solved analytically, the plot of the Eq. (\ref{hor}) is depicted in Fig. \ref{fig:ep} for different values of CoS parameter ($b$), and spin parameter ($a$) with fixed value of magnetic monopole charge ($g=0.1$).


\subsection{Static limit Surface and Ergo-region}

\begin{figure*}[ht]
\begin{tabular}{c c c c}
\includegraphics[width=.21\linewidth]{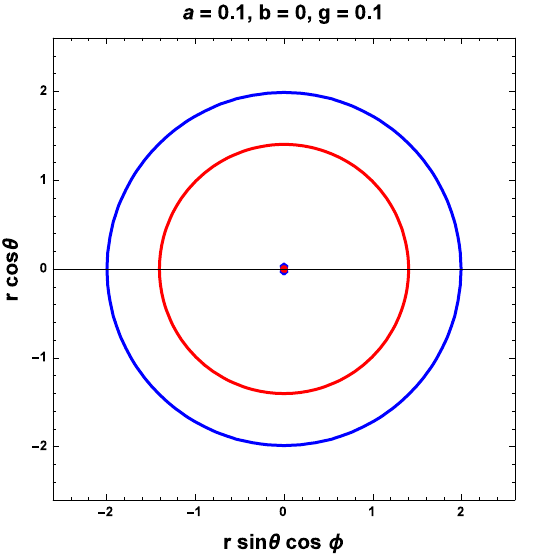}
\includegraphics[width=.21\linewidth]{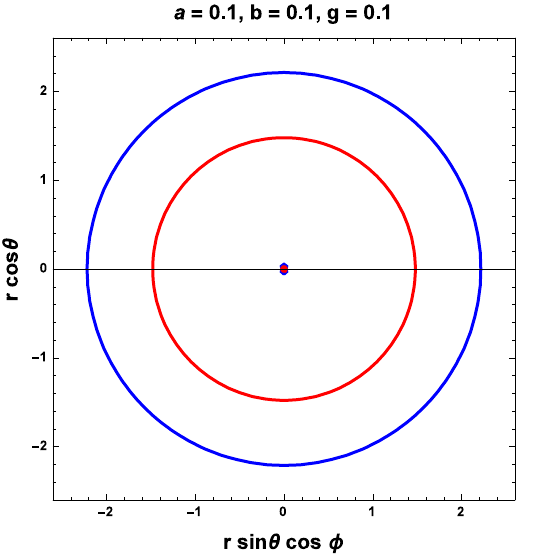}
\includegraphics[width=.21\linewidth]{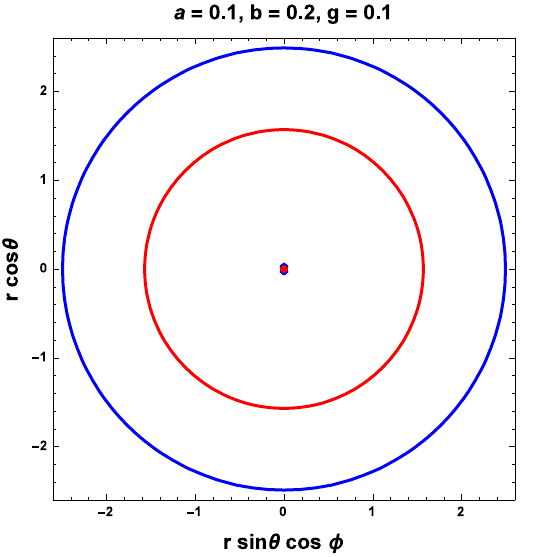}
\includegraphics[width=.21\linewidth]{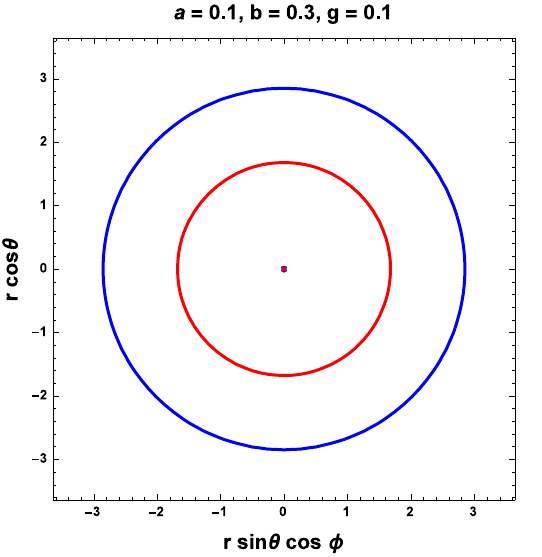}\\
\includegraphics[width=.21\linewidth]{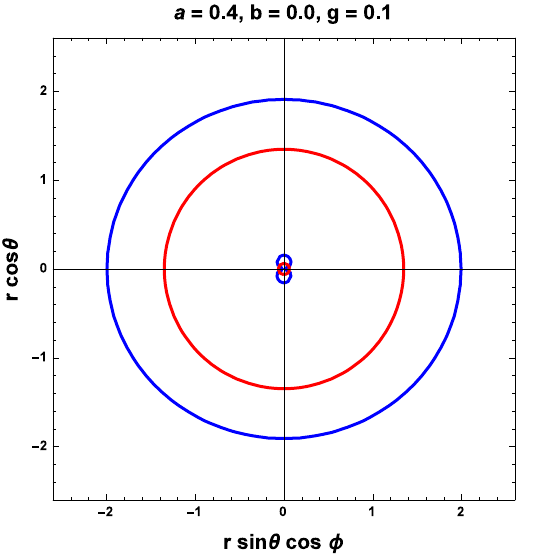}
\includegraphics[width=.21\linewidth]{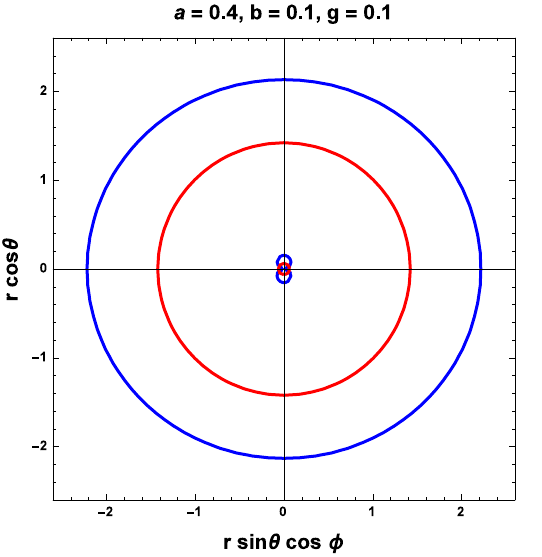}
\includegraphics[width=.21\linewidth]{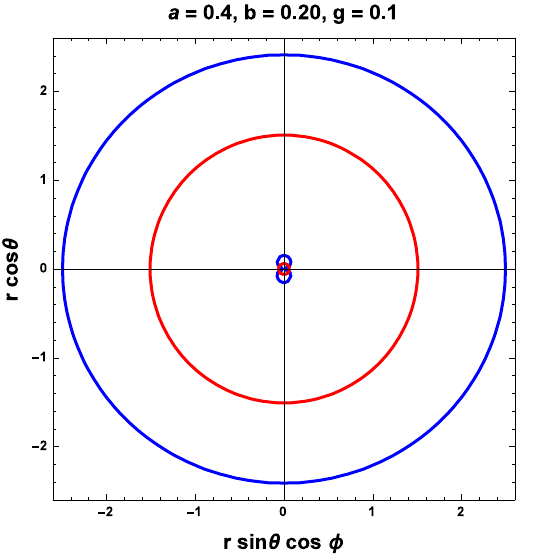}
\includegraphics[width=.21\linewidth]{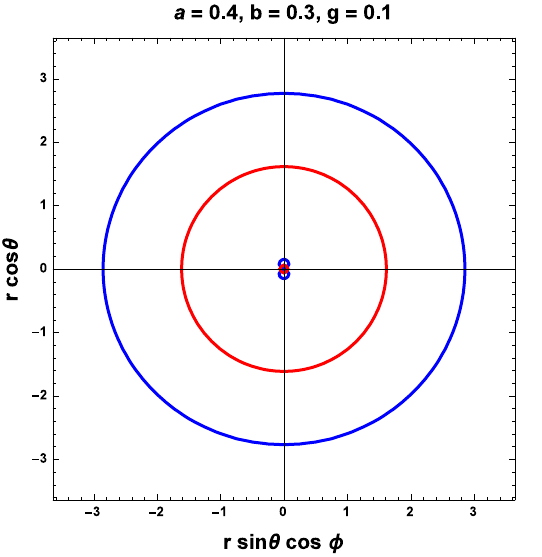}\\
\includegraphics[width=.21\linewidth]{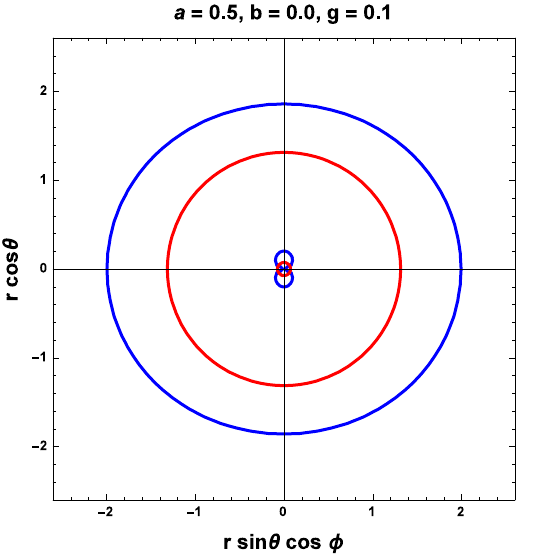}
\includegraphics[width=.21\linewidth]{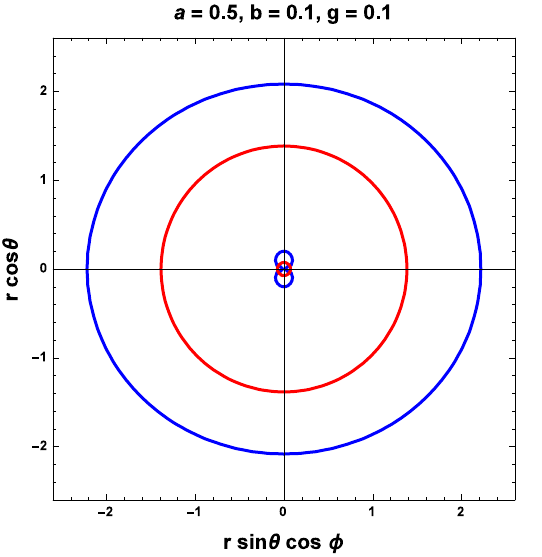}
\includegraphics[width=.21\linewidth]{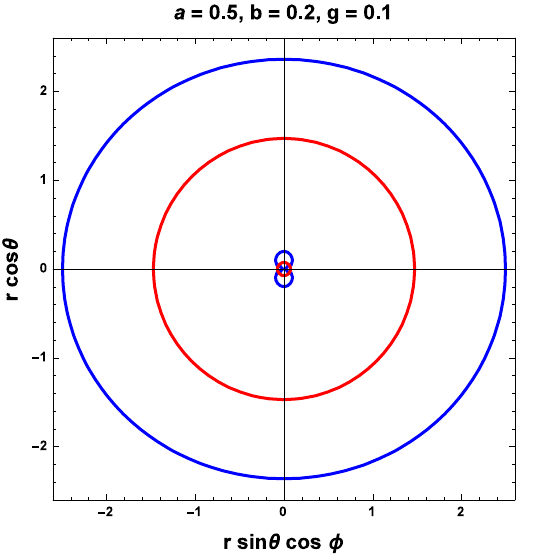}
\includegraphics[width=.21\linewidth]{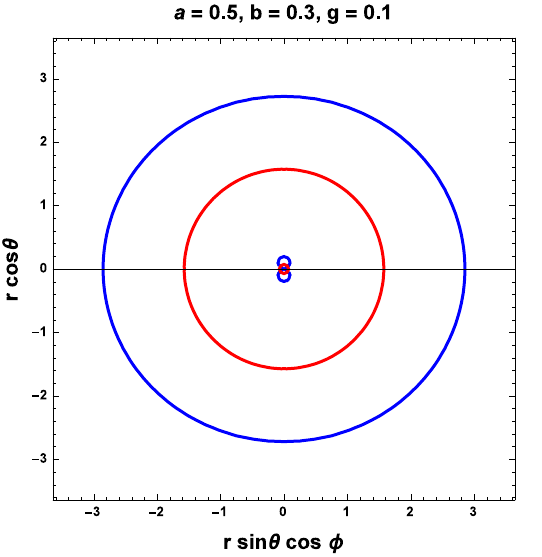}\\
\includegraphics[width=.21\linewidth]{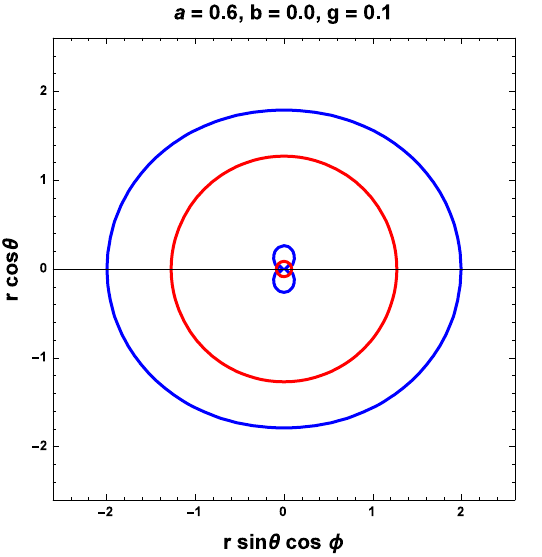}
\includegraphics[width=.21\linewidth]{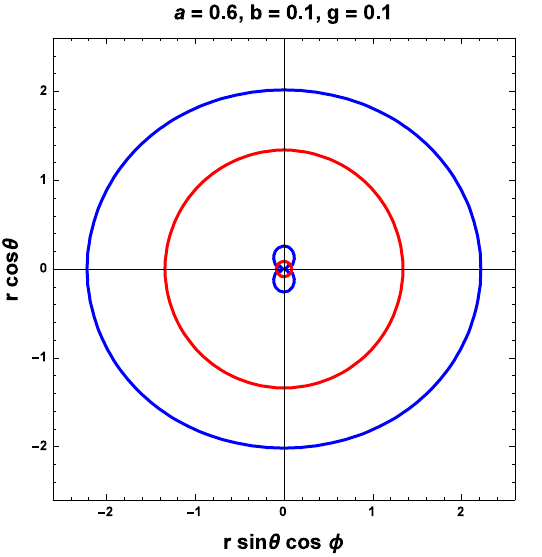}
\includegraphics[width=.21\linewidth]{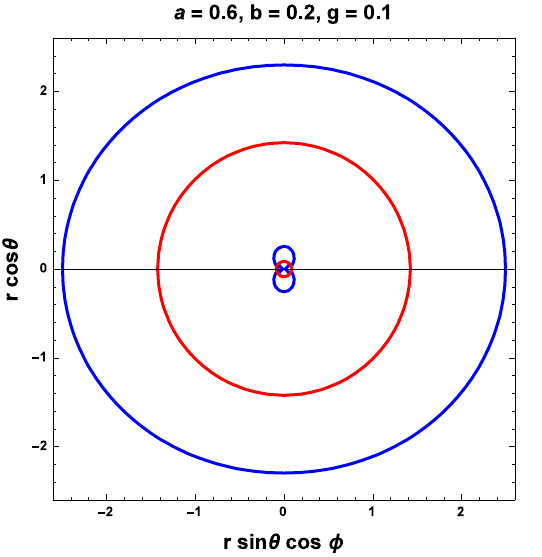}
\includegraphics[width=.21\linewidth]{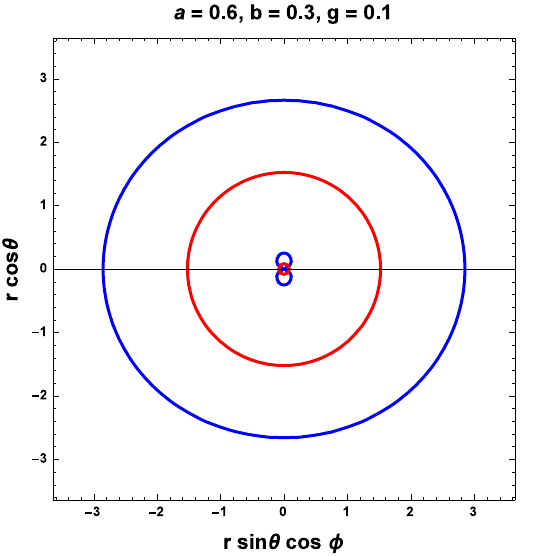}
\end{tabular}
\caption{Plot of ergo-region in x-z plane for different value of CoS parameter with a fixed value of the spin parameter and magnetic monopole charge. }
\label{fig:er}
\end{figure*}

The static limit surface is the region between the event horizon and the ergo-region and it is defined as a surface, where no observer can be at rest and static. We plot the ergo-region in the $x-z$ plane as depicted in Fig. \ref{fig:er}, for different values of black hole parameters, $(a,b,g)$.
The static limit  is defined by, ($g_{tt}=0$)

\begin{table}[ht]
 \begin{center}
 \begin{tabular}{ |l | l   | l   | l   |  l |  l | l | l | l | l | }
\hline
 \multicolumn{1}{|c|}{ $b$} &\multicolumn{1}{c|}{$r_1$}  &\multicolumn{1}{c|}{$r_2$}  &\multicolumn{1}{c|}{$\delta$} &\multicolumn{1}{c|}{$b$}&\multicolumn{1}{c|}{$r_1$}&\multicolumn{1}{c|}{$r_2$}&\multicolumn{1}{c|}{$\delta$}\\
            \hline
\multicolumn{1}{|c|}{ $$} &\multicolumn{1}{c}{$a=0.0,$}  &\multicolumn{1}{c}{$g=0.1$}  &\multicolumn{1}{c|}{$$} &\multicolumn{1}{c|}{$$}&\multicolumn{1}{c}{$a=0.1,$}&\multicolumn{1}{c}{$g=0.1$}&\multicolumn{1}{c|}{$$}\\
            \hline
 
            \,\,\,\,\,0.10 ~~  &~~0.556~~  & ~~2.210~~ & ~~2.154~~ & ~~0.0~~& ~~0.965~~& ~~1.752~~& ~~0.787~~ \\            
                     \hline
              \,\,\,\,\,0.10 ~~  &~~0.159~~  & ~~2.153~~ & ~~1.994~~ & ~~0.10~~& ~~0.774~~& ~~2.301~~& ~~1.577~~ \\         
\hline
\,\,\,\,\,0.20 ~~  &~~0.324~~  & ~~2.028~~ & ~~1.704~~ & ~~0.20~~& ~~0.656~~& ~~2.895~~& ~~2.239~~ \\ 
\hline
\,\,\,\,\,0.30 ~~  &~~0.607~~  & ~~1.788~~ & ~~1.181~~ & ~~0.30~~& ~~0.564~~& ~~3.560~~& ~~2.996~~ \\ 
\hline
\multicolumn{1}{|c|}{ $$} &\multicolumn{1}{c}{$a=0.3,$}  &\multicolumn{1}{c}{$ g=0.1$}  &\multicolumn{1}{c|}{$$} &\multicolumn{1}{c|}{$$}&\multicolumn{1}{c}{$a=0.5,$}&\multicolumn{1}{c}{$g=0.1$}&\multicolumn{1}{c|}{$$}\\
            \hline
  \,\,\,\,\,0.0 ~~  &~~0.116~~  & ~~2.169~~ & ~~2.053~~ & ~~0.0~~& ~~1.275~~& ~~1.510~~& ~~0.230~~ \\            
                     \hline
              \,\,\,\,\,0.10 ~~  &~~0.264~~  & ~~2.109~~ & ~~1.845~~ & ~~0.10~~& ~~0.940~~& ~~2.212~~& ~~1.272~~ \\         
\hline
\,\,\,\,\,0.20 ~~  &~~0.453~~  & ~~1.973~~ & ~~1.520 & ~~0.20~~& ~~0.807~~& ~~2.832~~& ~~2.025~~ \\ 
\hline
\,\,\,\,\,0.30 ~~  &~~0.769~~  & ~~1.696~~ & ~~0.927 & ~~0.30~~& ~~0.717~~& ~~3.602~~& ~~2.887~~ \\ 
\hline
\multicolumn{1}{|c|}{ $$} &\multicolumn{1}{c}{$a=0.7,$}  &\multicolumn{1}{c}{$ g=0.1$}  &\multicolumn{1}{c|}{$$} &\multicolumn{1}{c|}{$$}&\multicolumn{1}{c}{$a=0.9,$}&\multicolumn{1}{c}{$g=0.1$}&\multicolumn{1}{c|}{$$}\\
            \hline
  \,\,\,\,\,0.0 ~~  &~~0.199~~  & ~~2.081~~ & ~~1.882~~ & ~~0.0~~& ~~---~~& ~~---~~& ~~---~~ \\            
                     \hline
              \,\,\,\,\,0.10 ~~  &~~0.388~~  & ~~2.012~~ & ~~1.624~~ & ~~0.10~~& ~~1.268~~& ~~1.966~~& ~~0.698~~ \\         
\hline
\,\,\,\,\,0.20 ~~  &~~0.631~~  & ~~1.848~~ & ~~1.217~~ & ~~0.20~~& ~~1.039~~& ~~2.688~~& ~~1.647~~ \\ 
\hline
\,\,\,\,\,0.30 ~~  &~~1.193~~  & ~~1.340~~ & ~~0.147 & ~~0.30~~& ~~0.921~~& ~~3.488~~& ~~2.567~~ \\ 
\hline
        \end{tabular}
        \caption{The  SLS  of  Letelier-Bardeen-like black hole for different values of magnetic monopole charge ($g$) and CoS parameter ($b$), where $\delta=r_2-r_1$.}
\label{tab:sls}
    \end{center}
\end{table}

The numerical values of SLS are tabulated in Tab. \ref{tab:sls}, for a given value of spin parameter and angle  $(a,\theta)$, and for different values of CoS parameter ($b$). The SLS has no root when ($b>b_s$) and has two simple zeros, if ($b< b_s$). Similarly, we can also see the effect spin parameter on SLS with a fixed value of the CoS parameter and magnetic monopole charge. The size of the SLS increases with 
the increase in the spin parameter ($a$). We also notice that the size of the SLS decreases with increases in the CoS parameter and increases with increases in the spin parameter. Thus the the effect of CoS parameter and spin parameter on SLS are opposite to each other.

We also investigate the effect of black hole parameters ($a,b,g$) on ergo-region which is plotted in Fig. \ref{fig:er}. We plot the ergo-region for different values of the CoS parameter ($b$) with for fixed values of ($a,g$). We notice that when we increase the value of the CoS parameter ($b$) the resulting ergo-region also increases.

\section{Geodesics around the Letelier-Bardeen-Kerr black hole}
Let us consider the motion of the massless particles (photons) moving in the space-time (\ref{bhs}). We shall be interested in the photon motion in the equatorial plane by restricting $\theta=\pi/2$. The corresponding equations of motion can be obtained using the Hamiltonian-Jacobi formalism, \cite{Belhaj:2020okh,Belhaj:2022kek}.
 
The equations of motion are obtained as,
\begin{eqnarray}
       \Sigma\frac{d r}{d\tau}&=& \sqrt{\mathcal{R}(r)} \\
      \Sigma\frac{d\theta}{d\tau}&=& \sqrt{\Theta(\theta)}
\end{eqnarray}
 where $\mathcal{R}(r)$ and ${\Theta}(\theta)$ are given below as following
     \begin{eqnarray}
         \mathcal{R}(r)&&=\left(E(r^{2}+a^{2})-a~L\right)^{2}-\Delta\left(\mathcal{\kappa}+(L-a~E)^{2}\right)\\
         {\Theta}(\theta)&&= \mathcal{\kappa}-\cos^{2}\theta(\left(a^{2}-E^{2}+\frac{L^{2}}{\sin^{2}\theta}\right)
         \label{shadoweqns}
     \end{eqnarray}
where $E$ and $L$ are the energy and angular momentum of the particle respectively.   

The radial equation can be put in the form, $\dot r^2+V_{eff}(r)= E^2$, where $V_{eff}$ is the effective potential, given by 
\begin{equation}
V_{eff}=\frac{E^2 r^4+\Delta(L-aE)^2-[(r^2+a^2)E-aL]^2}{r^4}
\label{eq:ep}
\end{equation}

The  null circular geodesics  obey the conditions $V_{eff}=0, \,\, {\partial V_{eff}}/{\partial r}=0$ and ${\partial^2 V_{eff}}/{\partial r^2}>0$, which gives
\begin{equation}
\left((1-b)-\frac{3Mr^4_p}{(g^2+r^2_p)^{5/2}}\right)^2-\frac{4a^2M(-2g^2+r^2_p)}{(g^2+r^2_p)^{5/2}}=0.
\label{rp}
\end{equation}

\begin{table}[ht]
 \begin{center}
 \begin{tabular}{ |l | l   | l   | l   |  l |  l | l | l | l | l | l |}
\hline
  \multicolumn{1}{|c}{ } &\multicolumn{1}{c}{}  &\multicolumn{1}{c}{}  &\multicolumn{1}{c }{ }&\multicolumn{1}{c }{ }&\multicolumn{1}{c }{ $r_p$}&\multicolumn{1}{c }{ }&\multicolumn{1}{c }{ }&\multicolumn{1}{c }{ } &\multicolumn{1}{c}{}&\multicolumn{1}{c|}{}\\
            \hline
  \multicolumn{1}{|c}{ $$} &\multicolumn{1}{c}{$a=0.1$}  &\multicolumn{1}{c|}{$$}  &\multicolumn{1}{c}{$a=0.3$} &\multicolumn{1}{c|}{$$}&\multicolumn{1}{c}{$a=0.5$}&\multicolumn{1}{c|}{}&\multicolumn{1}{c}{$a=0.7$}&\multicolumn{1}{c|}{$$}&\multicolumn{1}{c}{$a=0.9$}&\multicolumn{1}{c|}{}\\
               \hline
   \multicolumn{1}{|c|}{ $\bf g$} &\multicolumn{1}{c|}{$\bf r_{p1}$}  &\multicolumn{1}{c|}{$\bf r_{p2}$}  &\multicolumn{1}{c|}{$ \bf r_{p1}$} &\multicolumn{1}{c|}{$\bf r_{p2}$}&\multicolumn{1}{c|}{$\bf r_{p1}$}&\multicolumn{1}{c|}{$\bf r_{p2}$}&\multicolumn{1}{c|}{$\bf r_{p1}$}&\multicolumn{1}{c|}{$\bf r_{p2}$}&\multicolumn{1}{c|}{$\bf r_{p1}$}&\multicolumn{1}{c|}{$\bf r_{p2}$}\\
            \hline
            \,\,\,\,\,0.1 ~~  &~~2.912~~  & ~~3.302~~ & ~~2.563~~ & ~~3.940~~& ~~2.281~~& ~~4.10~~& ~~2.007~~& ~~4.231~~& ~~1.700~~ &~~4.339~~\\            
                     \hline
                     \,\,\,\,\,0.2 ~~  &~~2.884~~  & ~~3.672~~ & ~~2.526~~ & ~~3.922~~& ~~2.235~~& ~~4.086~~& ~~1.945~~& ~~4.215~~& ~~1.594~~ &~~4.324~~\\            
                     \hline
                     \,\,\,\,\,0.3 ~~  &~~2.834~~  & ~~3.640~~ & ~~2.462~~ & ~~3.893~~& ~~2.151~~& ~~4.058~~& ~~1.822~~& ~~4.188~~& ~~0.490~~ &~~4.298~~\\            
                     \hline
                     \,\,\,\,\,0.4 ~~  &~~2.761~~  & ~~3.592~~ & ~~2.364~~ & ~~3.850~~& ~~2.053~~& ~~4.018~~& ~~1.565~~& ~~4.150~~& ~~0.642~~ &~~4.262~~\\            
                     \hline
                     \,\,\,\,\,0.5 ~~  &~~2.657~~  & ~~3.529~~ & ~~2.215~~ & ~~3.793~~& ~~1.763~~& ~~3.965~~& ~~0.813~~& ~~4.100~~& ~~0.770~~ &~~4.213~~\\            
                     \hline
                     \,\,\,\,\,0.6 ~~  &~~2.512~~  & ~~3.447~~ & ~~1.969~~ & ~~3.721~~& ~~0.966~~& ~~3.898~~& ~~0.907~~& ~~4.036~~& ~~0.888~~ &~~4.152~~\\            
                     \hline
                     \,\,\,\,\,0.7 ~~  &~~2.293~~  & ~~3.342~~ & ~~---~~ & ~~3.629~~& ~~1.027~~& ~~3.813~~& ~~1.013~~& ~~3.956~~& ~~1.007~~ &~~4.076~~\\            
                     \hline
                     \,\,\,\,\,0.8 ~~  &~~1.790~~  & ~~3.206~~ & ~~---~~ & ~~3.515~~& ~~1.136~~& ~~3.709~~& ~~1.134~~& ~~3.858~~& ~~1.133~~ &~~3.985~~\\            
                     \hline
                     \,\,\,\,\,0.9 ~~  &~~---~~  & ~~3.062~~ & ~~---~~ & ~~3.368~~& ~~1.276~~& ~~3.756~~& ~~1.275~~& ~~3.737~~& ~~1.274~~ &~~3.869~~\\  
                   \hline
        \end{tabular}
        \caption{The numerical values of photon radius for different values of magnetic monopole charge ($g$) and spin parameter ($a$) with a fixed value of CoS parameter ($b$).}
\label{tr3}
    \end{center}
\end{table}

In Tab. \ref{tr3} and Tab. \ref{tr2}, we can see that the photon radius ($r_p$) of the obtained black hole solution increases with increases in the spin parameter and CoS parameter. The photon radius decreases with increased magnetic monopole charge ($g$). We can say that the effect of magnetic monopole charge is opposite with spin parameter and CoS parameter.

\begin{table}[ht]
 \begin{center}
 \begin{tabular}{ |l | l   | l   | l   |  l |  l | l | l | l | l | l |}
\hline
  \multicolumn{1}{|c}{ } &\multicolumn{1}{c}{}  &\multicolumn{1}{c}{}  &\multicolumn{1}{c }{ }&\multicolumn{1}{c }{ }&\multicolumn{1}{c }{ $r_p$}&\multicolumn{1}{c }{ }&\multicolumn{1}{c }{ }&\multicolumn{1}{c }{ } &\multicolumn{1}{c}{}&\multicolumn{1}{c|}{}\\
            \hline
  \multicolumn{1}{|c}{ $$} &\multicolumn{1}{c}{$a=0.1$}  &\multicolumn{1}{c|}{$$}  &\multicolumn{1}{c}{$a=0.3$} &\multicolumn{1}{c|}{$$}&\multicolumn{1}{c}{$a=0.5$}&\multicolumn{1}{c|}{}&\multicolumn{1}{c}{$a=0.7$}&\multicolumn{1}{c|}{$$}&\multicolumn{1}{c}{$a=0.9$}&\multicolumn{1}{c|}{}\\
               \hline
   \multicolumn{1}{|c|}{ $\bf b$} &\multicolumn{1}{c|}{$\bf r_{p1}$}  &\multicolumn{1}{c|}{$\bf r_{p2}$}  &\multicolumn{1}{c|}{$ \bf r_{p1}$} &\multicolumn{1}{c|}{$\bf r_{p2}$}&\multicolumn{1}{c|}{$\bf r_{p1}$}&\multicolumn{1}{c|}{$\bf r_{p2}$}&\multicolumn{1}{c|}{$\bf r_{p1}$}&\multicolumn{1}{c|}{$\bf r_{p2}$}&\multicolumn{1}{c|}{$\bf r_{p1}$}&\multicolumn{1}{c|}{$\bf r_{p2}$}\\
            \hline
            \,\,\,\,\,0.1 ~~  &~~1.286~~  & ~~3.062~~ & ~~1.278~~ & ~~3.368~~& ~~1.276~~& ~~3.578~~& ~~1.257~~& ~~3.737~~& ~~1.274~~ &~~3.869~~\\            
                     \hline
                     \,\,\,\,\,0.2 ~~  &~~2.146~~  & ~~3.578~~ & ~~1.280~~ & ~~3.909~~& ~~1.278~~& ~~4.118~~& ~~1.277~~& ~~4.280~~& ~~1.276~~ &~~4.414~~\\            
                     \hline
                     \,\,\,\,\,0.3 ~~  &~~3.059~~  & ~~4.233~~ & ~~2.209~~ & ~~4.566~~& ~~1.344~~& ~~3.781~~& ~~1.315~~& ~~4.947~~& ~~1.303~~ &~~5.087~~\\            
                     \hline
                     \,\,\,\,\,0.4 ~~  &~~3.927~~  & ~~5.061~~ & ~~3.358~~ & ~~5.407~~& ~~2.802~~& ~~5.632~~& ~~1.494~~& ~~5.808~~& ~~1.388~~ &~~5.956~~\\            
                     \hline
                     \,\,\,\,\,0.5 ~~  &~~5.018~~  & ~~6.178~~ & ~~4.494~~ & ~~6.546~~& ~~4.070~~& ~~6.787~~& ~~3.653~~& ~~6.977~~& ~~3.173~~ &~~7.138~~\\            
                     \hline
                     \,\,\,\,\,0.6 ~~  &~~6.867~~  & ~~7.805~~ & ~~6.039~~ & ~~8.009~~& ~~5.636~~& ~~8.476~~& ~~5.275~~& ~~8.687~~& ~~4.928~~ &~~8.866~~\\            
                     \hline
                     \,\,\,\,\,0.7 ~~  &~~9.073~~  & ~~10.45~~ & ~~8.504~~ & ~~10.920~~& ~~8.087~~& ~~11.78~~& ~~7.721~~& ~~11.471~~& ~~7.388~~ &~~11.67~~\\            
                     \hline
                     \,\,\,\,\,0.8 ~~  &~~14.09~~  & ~~15.66~~ & ~~13.34~~ & ~~16.23~~& ~~12.80~~& ~~16.31~~& ~~12.46~~& ~~16.91~~& ~~12.101~~ &~~17.17~~\\            
                     \hline
                     \,\,\,\,\,0.9 ~~  &~~28.74~~  & ~~31.06~~ & ~~27.85~~ & ~~31.87~~& ~~27.21~~& ~~32.42~~& ~~26.89~~& ~~32.85~~& ~~26.21~~ &~~33.23\\            
                     \hline
        \end{tabular}
        \caption{The numerical values of photon radius for different values of CoS parameter ($b$) and spin parameter ($a$) with fixed value of magnetic monopole charge ($g$).}
\label{tr2}
    \end{center}
\end{table}

We obtain the critical values of the impact factor by solving the equations ($\ref{shadoweqns}$) for null geodesics. Using the boundary conditions at $\mathcal{R}(r)=0$ and ${d\mathcal{R}}/{d r}$ at $r={r}_p$, we get
 
\begin{eqnarray}
    &&\eta=\frac{r^{2}(16a^{2}\Delta-16\Delta^{2}+8r\Delta\Delta_r-r^{2}\Delta_r^{2})}{a^{2}\Delta_r^{2}},\nonumber\\
    && \xi=\frac{(r^{2}+a^{2})\Delta_r-4r\Delta_r}{a\Delta_r}
    \label{eq:im}
\end{eqnarray}
where, $\xi=L/E$ and $\eta= {\cal K}/E^2$ are the two dimensionless impact parameters and $\Delta_r$ is first derivative of $\Delta(r)$ with respect to $r$. The impact parameter (\ref{eq:im}) reduces to the impact factor of rotating Bardeen black hole in the absence of CoS parameter $(b=0)$, rotating Letelier black hole when ($g=0$) as well as  Kerr black hole in the limit of  $(g=b=0)$. Further, for an observer at the equatorial plane ($\theta=\pi/2$), it simplifies to
\begin{eqnarray}
  x=-\xi,\qquad\qquad \text{and}\qquad\qquad   y=\pm \sqrt{\eta}
\end{eqnarray}

\begin{figure*}[ht]
\begin{tabular}{c c c c}
\includegraphics[width=.30\linewidth]{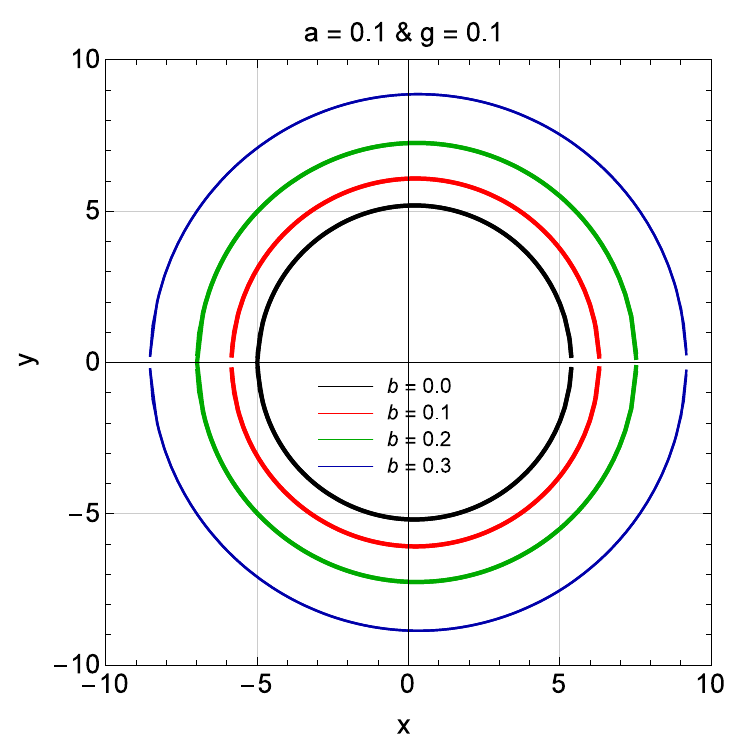}
\includegraphics[width=.30\linewidth]{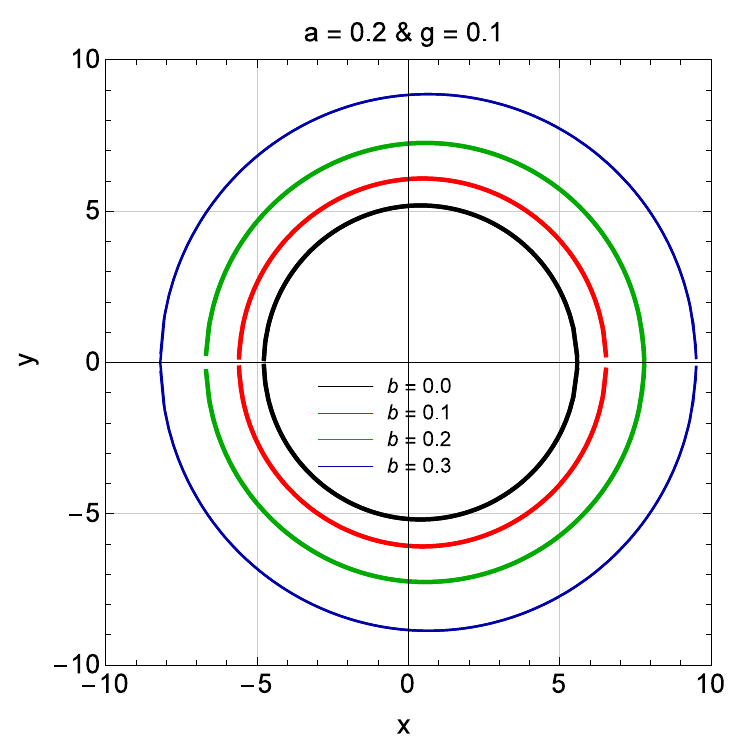}
\includegraphics[width=.30\linewidth]{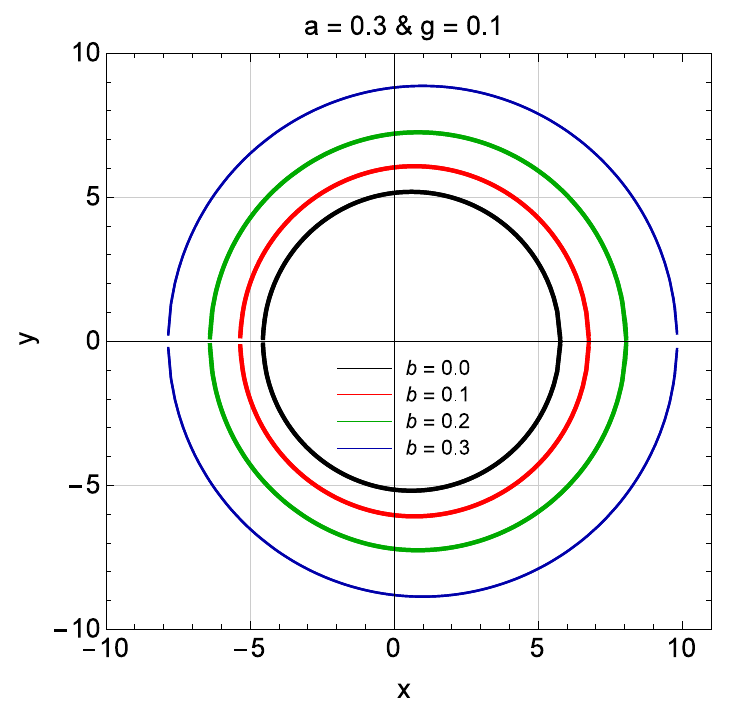}\\
\includegraphics[width=.30\linewidth]{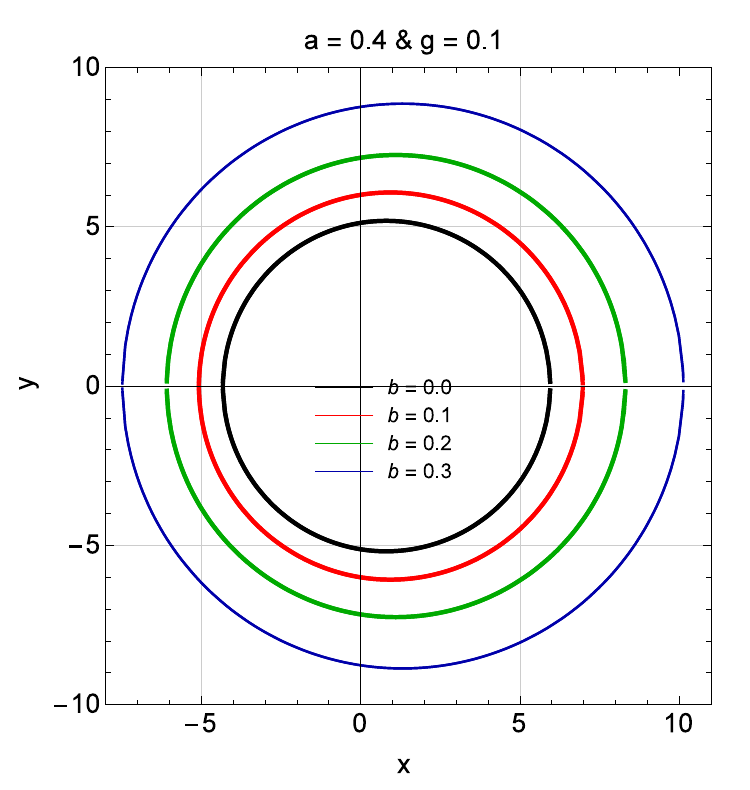}
\includegraphics[width=.30\linewidth]{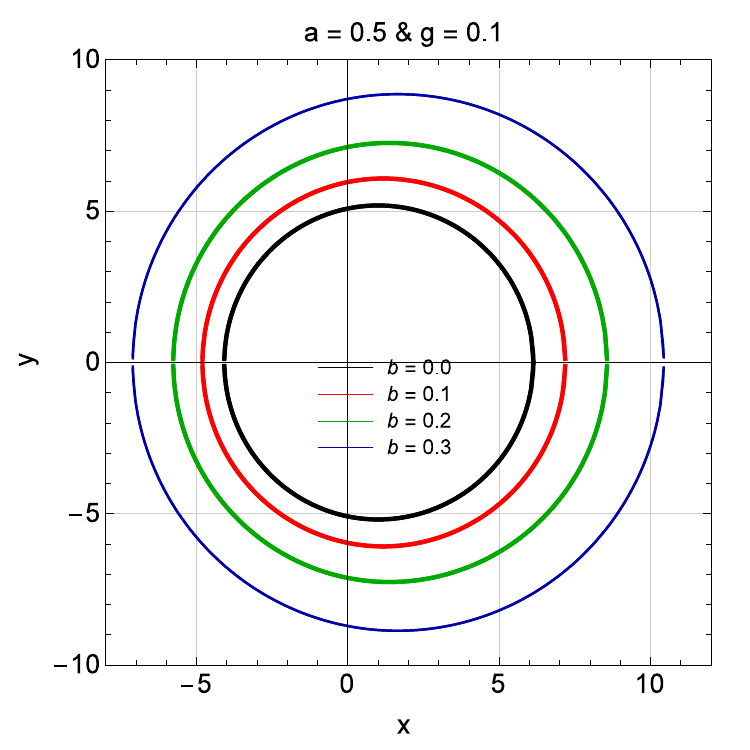}
\includegraphics[width=.30\linewidth]{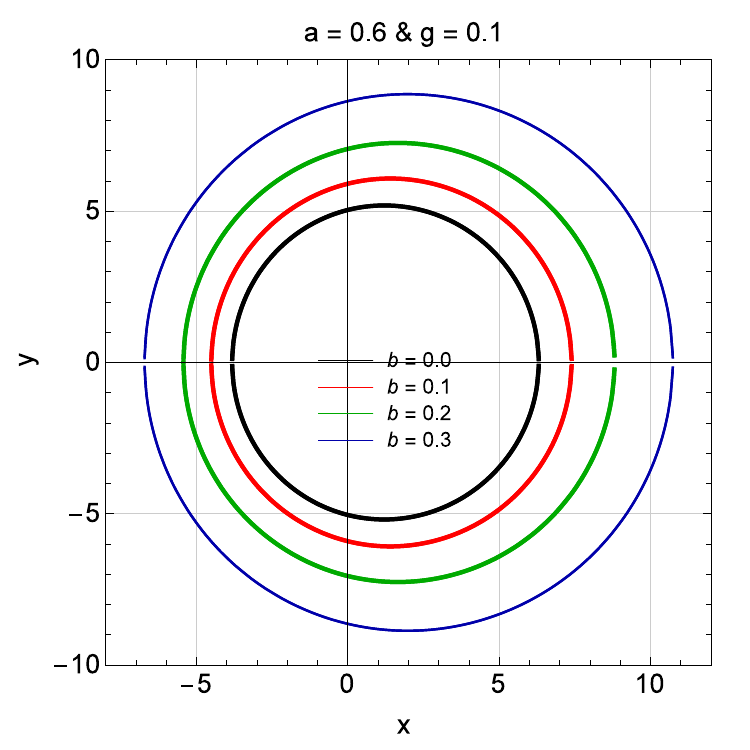}\\
\includegraphics[width=.30\linewidth]{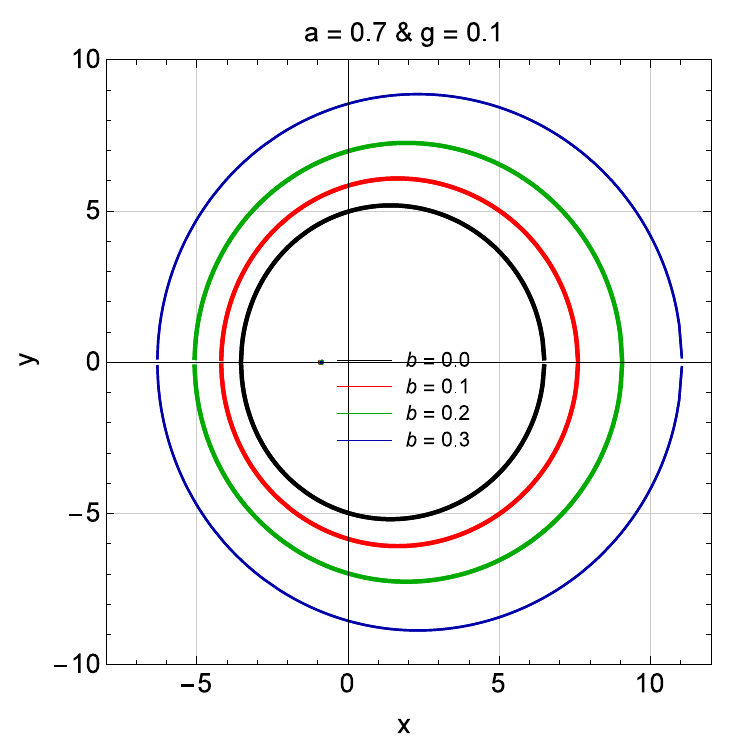}
\includegraphics[width=.30\linewidth]{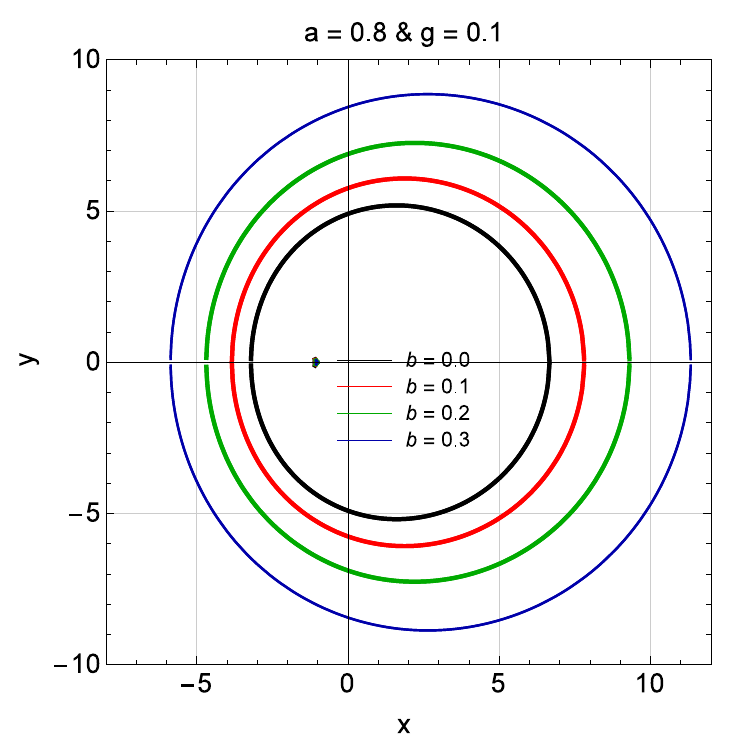}
\includegraphics[width=.30\linewidth]{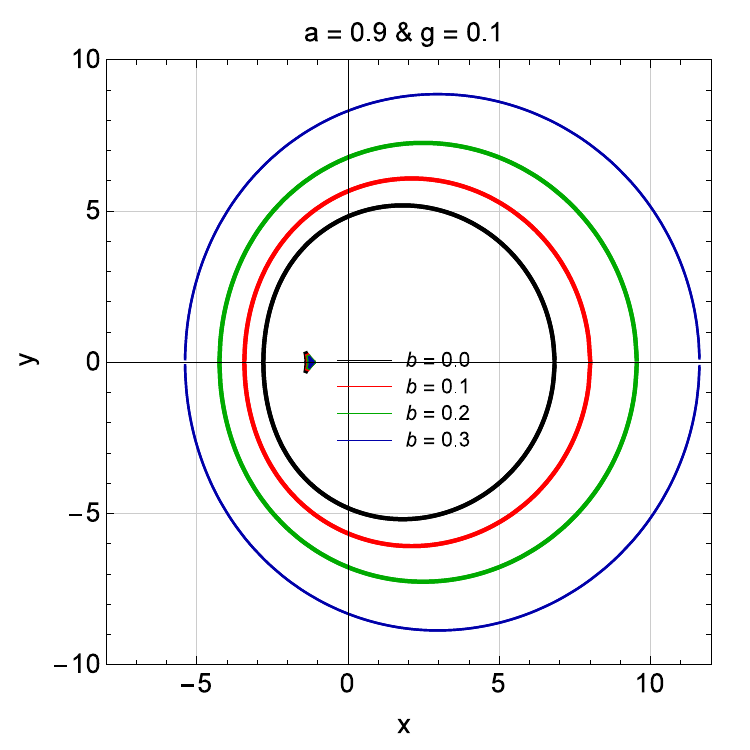}
\end{tabular}
\caption{The plot of shadow for different values of CoS parameter ($b$) with a fixed value of the spin parameter ($a$), magnetic monopole charge ($g$), and mass of the black hole ($M=1$).}
\label{eq:ep}
\end{figure*}

 \begin{figure*}[ht]
\begin{tabular}{c c c c}
\includegraphics[width=.55\linewidth]{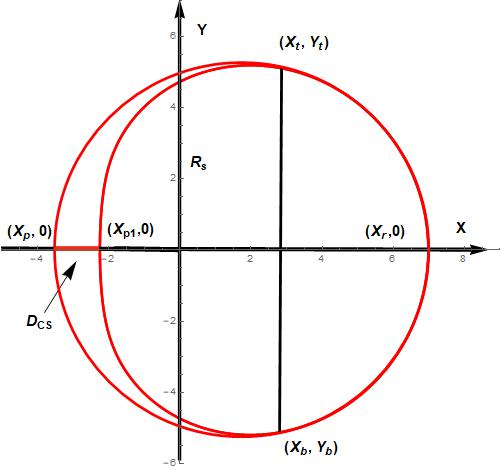}
\end{tabular}
\caption{Schematic representation of the observables for the shadow of rotating black holes.}
\label{fig:schr}
\end{figure*}

The shadow image of the obtained black hole solutions (\ref{bhs}) for choice of parameters ($a,b,g$) and $\theta_0$ are plotted in Fig. \ref{eq:ep}. In Fig. \ref{eq:ep},  we can see that the size of shadow images will increase the CoS parameter ($b$) and ($a$). The distortion of the shadow images arises at the higher values of the spin parameter ($a$).

\section{Summary and Results}
We have investigated the ergo-regions and shadows of the Bardeen-Kerr black hole in a cloud of strings in the analysis so far. The geodesic equation of the photons is obtained in this geometry and the shadows are plotted for different values of the parameters $a$ and $b$. The shadow deformation can be used to determine the spin, $a$ of the black hole for different values of the CoS parameter, $b$ as explained below.

\begin{figure*}[ht]
\begin{tabular}{c c c c}
\includegraphics[width=.50\linewidth]{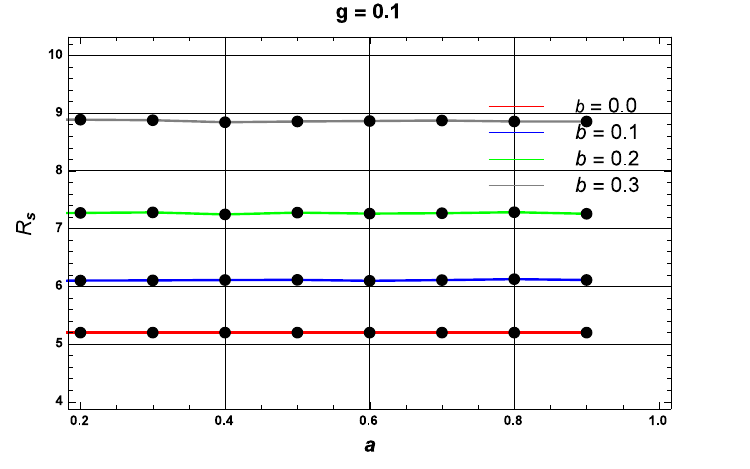}
\includegraphics[width=.50\linewidth]{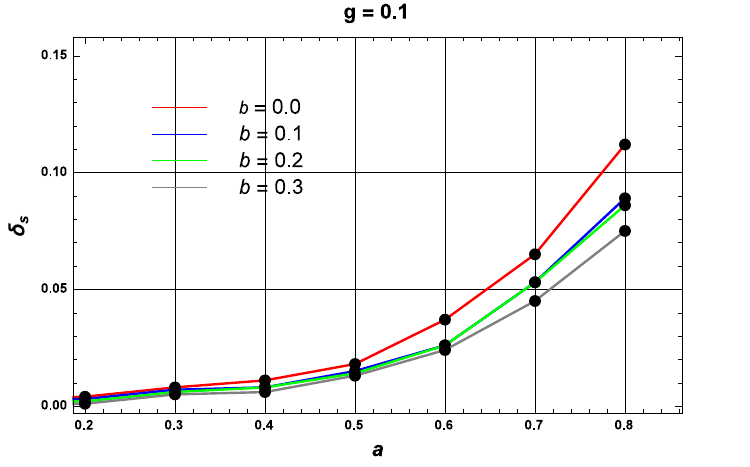}
\end{tabular}
\caption{The plot of $R_s$ and $\delta_s$ vs $a$ for different values of CoS parameter with a fixed value of the magnetic charge. }
\label{fig:sh1}
\end{figure*}

The relation between the celestial coordinate ($x,y$) and impact parameter ($\eta, \xi$) is given in Eq. (27). The size and shape of the black hole shadow solution depend upon the parameters $(a,b,g)$ and inclination angle $\theta$. We can use the Schmidt representation of the rotating black hole as depicted in Fig. \ref{fig:schr} to determine the size of the black hole shadow using the following relation \cite{72}

\begin{equation}
R_s=\frac{(x_t-x_r)^2+Y_t^2}{2|x_t-x_r|}.
\label{d1}
\end{equation}
and  the  distortion parameter  defined as  a ratio  of  $D_s$ and $R_s$  reads as
\begin{equation}
\delta=\frac{D_s}{R_s}=\frac{|x_p-x'_P|}{R_s}.
\label{d2}
\end{equation}
where, $r,l, t$, and $b$, refer to the right, left, top, and bottom of the shadow boundary, and ($x_P,0$) and ($x_{p1},0$) are the points where the shadow cut the horizontal axis at the opposite side of  $(x_r, 0)$ (see Fig. \ref{fig:sh1}).

The black hole’s shadow $(R_s)$ and distortion parameter $(\delta_s)$ are plotted in the Fig. \ref{fig:sh1}. We can see that the shadow radius ($R_s$) increases with the CoS parameter ($b$) and is approximately constant with the spin parameter ($a$). The distortion $\delta_s$ of the shadow image increases with the spin parameter and decreases with the CoS parameter. The measurement of the distortion parameter, $(\delta_s)$ can be used to determine black hole spin from the above figure by parametric fitting. It would be interesting to compare our results with observational data and see the signatures of black holes created in the early universe, if any.


\section*{Data Availability Statement} 
Data sharing is not applicable to this article as no experimental data were used or analyzed during the current study.

\begin{acknowledgements}  
  The work of BKV is supported by a UGC fellowship. DVS thanks to the DST-SERB project (grant no. EEQ/2022/000824) under EEQ scheme.
\end{acknowledgements}



\begin{thebibliography}{99}
  \bibitem{frad}E. S. Fradkin and A. A. Tseytlin, Phys. Lett. B 163 
(1985)  123.

\bibitem{Ghosh:2018bxg}
S.~G.~Ghosh, D.~V.~Singh and S.~D.~Maharaj,
Phys. Rev. D \textbf{97} (2018) no.10, 104050.
\bibitem{Singh:2022xgi}
D.~V.~Singh, S.~G.~Ghosh and S.~D.~Maharaj,
Nucl. Phys. B \textbf{981} (2022), 115854.


\bibitem{Singh:2017bwj}
D.~V.~Singh, M.~S.~Ali and S.~G.~Ghosh,
Int. J. Mod. Phys. D \textbf{27} (2018) no.12, 1850108.

\bibitem{Maluf:2018lyu} 
  R.~V.~Maluf and J.~C.~S.~Neves,
  Phys.\ Rev.\ D {\bf 97}, 104015 (2018).
\bibitem{bambi}
C. Bambi and L. Modesto, Phys. Lett. B {\bf 721}, 329 (2013).
\bibitem{abg11}
 E. Ayon-Beato and A. Garcia,  Gen. Rel. Grav. \textbf{31}, 629 (1999).
\bibitem{abg}
E. Ayon-Beato, A. Garcia, Phys. Lett. B \textbf{493} (2000) 149.
\bibitem{bard}
J. Bardeen, Proceedings of GR5 (Tiflis, U.S.S.R., 1968).

 \bibitem{dvs19}
D. V. Singh, S. G. Ghosh and S. D. Maharaj, Annals Phys.\  {\bf 412}, 168025 (2020).

  \bibitem{Singh:2020xju}
D.~V.~Singh and S.~Siwach,
Phys. Lett. B \textbf{808} (2020), 135658.
\bibitem{k5}
B. Narzilloev, J. Rayimbaev, S. Shaymatov,A. Abdujabbarov, B. Ahmedov, and C. Bambi ,
Phys. Rev. D \textbf{1} (2020) No.102, 104062.

\bibitem{Singh:2019wpu}
D.~V.~Singh, S.~G.~Ghosh and S.~D.~Maharaj,
Annals Phys. \textbf{412} (2020), 168025.

\bibitem{fr1}
S. Fernando, Int. Journal of Mod. Phys. D {\bf 26}, 1750071 (2017).
\bibitem{Singh:2017qur}
D.~V.~Singh and N.~K.~Singh,
Annals Phys. \textbf{383} (2017), 600-609.

\bibitem{Kumar:2018vsm}
A.~Kumar, D.~V Singh and S.~G.~Ghosh,
Eur. Phys. J. C \textbf{79} (2019) no.3, 275.

\bibitem{Singh:2020rnm}
B.~K.~Singh, R.~P.~Singh and D.~V.~Singh,
Eur. Phys. J. Plus \textbf{135} (2020) no.10, 862.
\bibitem{Yan:2023pxj}
Z.~Yan, X.~Zhang, M.~Wan and C.~Wu,
Eur. Phys. J. Plus \textbf{138} (2023) no.5, 377.
\bibitem{Singh:2022dth}
D.~V.~Singh, V.~K.~Bhardwaj and S.~Upadhyay,
Eur. Phys. J. Plus \textbf{137} (2022) no.8, 969.
\bibitem{Jafarzade:2020ova}
K.~Jafarzade, M.~Kord Zangeneh and F.~S.~N.~Lobo,
JCAP \textbf{04} (2021), 008.
\bibitem{Ghosh:2022gka}
R.~Ghosh, M.~Rahman and A.~K.~Mishra,
Eur. Phys. J. C \textbf{83} (2023), 91.
\bibitem{Belhaj:2020nqy}
A.~Belhaj, L.~Chakhchi, H.~El Moumni, J.~Khalloufi and K.~Masmar,
Int. J. Mod. Phys. A \textbf{35} (2020) No.27, 2050170.
\bibitem{l1}
P. S. Letelier, Phys. Rev. D \textbf{20}, 1294 (1979).
\bibitem{l2}
 P. S. Letelier, Nuovo Cim. B \textbf{63} 519 (1981).
 \bibitem{xu19}
 P. S. Letelier, Phys. Rev. D \textbf{28}, 2414 (1983).

 
\bibitem{Rodrigues:2022a}
M.~E.~Rodrigues and M.~V.~d.~S.~Silva,
Phys. Rev. D \textbf{106} (2022) no.8, 084016.

\bibitem{Rodrigues:2022b}
M.~E.~Rodrigues and H.~A.~Vieira,
Phys. Rev. D \textbf{106} (2022) no.8, 084015.

\bibitem{Belhaj:2022kek}
A.~Belhaj and Y.~Sekhmani,
Gen. Rel. Grav. \textbf{54} (2022) no.2, 17.
\bibitem{Singh:2020nwo}
D.~V.~Singh, S.~G.~Ghosh and S.~D.~Maharaj,
Phys. Dark Univ. \textbf{30} (2020), 100730.
\bibitem{Singh:2022ycn}
D.~V.~Singh, A.~Shukla and S.~Upadhyay,
Annals Phys. \textbf{447} (2022), 169157.
\bibitem{bkv}
B.~K.~Vishvakarma, D.~V.~Singh and S.~Siwach,
Eur. Phys. J. Plus \textbf{138} (2023) no.6, 536.

\bibitem{Hioki:2009na}
K.~Hioki and K.~Maeda,
Phys. Rev. D \textbf{80} (2009), 024042
\bibitem{EslamPanah:2020hoj}
B.~Eslam Panah, K.~Jafarzade and S.~H.~Hendi,
Nucl. Phys. B \textbf{961} (2020), 115269.
\bibitem{Sau:2022afl}
S.~Sau and J.~W.~Moffat,
Phys. Rev. D \textbf{107} (2023) no.12, 124003.

\bibitem{Belhaj:2020okh}
A.~Belhaj, H.~Belmahi, M.~Benali, W.~El Hadri, H.~El Moumni and E.~Torrente-Lujan,
Phys. Lett. B \textbf{812} (2021), 136025.
\bibitem{72}
V.~Perlick, O.~Y.~Tsupko and G.~S.~Bisnovatyi-Kogan,
Phys. Rev. D \textbf{92} (2015) no.10, 104031.

\bibitem{Kamruddin:2013iea}
A.~B.~Kamruddin and J.~Dexter,
Mon. Not. Roy. Astron. Soc. \textbf{434} (2013), 765.


\bibitem{Abdujabbarov:2016hnw}
A.~Abdujabbarov, M.~Amir, B.~Ahmedov and S.~G Ghosh
Phys. Rev. D \textbf{93} (2016) no.10, 104004.
\bibitem{Ghosh:2020ece}
S.~G.~Ghosh, M.~Amir and S.~D.~Maharaj,
Nucl. Phys. B \textbf{957} (2020), 115088.
\bibitem{Ahmed:2022qge}
F.~Ahmed, D.~V.~Singh and S.~G.~Ghosh,
Gen. Rel. Grav. \textbf{54} (2022) no.2, 21.
\bibitem{Kumar:2020owy}
R.~Kumar and S.~G.~Ghosh,
JCAP \textbf{07} (2020), 053.
\bibitem{Jana:2023sil}
S.~Jana and S.~Kar,
Phys. Rev. D \textbf{108} (2023) no.4, 044008.

\bibitem{Tan:2023ngk}
H.~S.~Tan,
Class. Quant. Grav. \textbf{40} (2023) no.19, 195010.

\bibitem{Kumar:2020hgm}
R.~Kumar, S.~G.~Ghosh and A.~Wang,
Phys. Rev. D \textbf{101} (2020) no.10, 104001.


\bibitem{Pantig:2022qak}
R.~C.~Pantig, A.~\"Ovg\"un and D.~Demir,
Eur. Phys. J. C \textbf{83} (2023) no.3, 250.
\bibitem{Psaltis:2018xkc}
D.~Psaltis,
Gen. Rel. Grav. \textbf{51} (2019) No.10, 137.
\bibitem{Moffat:2019uxp}
J.~W.~Moffat and V.~T.~Toth,
Phys. Rev. D \textbf{101} (2020) no.2, 024014.

\bibitem{Banerjee:2022iok}
I.~Banerjee, S.~Sau and S.~SenGupta,
JCAP \textbf{09} (2022), 066.


\bibitem{Afrin:2021imp}
M.~Afrin, R.~Kumar and S.~G.~Ghosh,
Mon. Not. Roy. Astron. Soc. \textbf{504} (2021), 5927-5940.
\bibitem{Kumar:2018ple}
R.~Kumar and S.~G.~Ghosh,
Astrophys. J. \textbf{892} (2020), 78.
\bibitem{Afrin:2023uzo}
M.~Afrin and S.~G.~Ghosh,
Mon. Not. Roy. Astron. Soc. \textbf{524} (2023) No.3, 3683-3691.
\bibitem{KumarWalia:2022aop}
R.~Kumar Walia, S.~G.~Ghosh and S.~D.~Maharaj,
Astrophys. J. \textbf{939} (2022) No.2, 77.
\bibitem{Afrin:2022ztr}
M.~Afrin, S.~Vagnozzi and S.~G.~Ghosh,
Astrophys. J. \textbf{944} (2023) No.2, 149.

\bibitem{Afrin:2021wlj}
M.~Afrin and S.~G.~Ghosh,
Astrophys. J. \textbf{932} (2022) No.1, 51.
\bibitem{Pulice:2023dqw}
B.~Pulice, R.~C.~Pantig, A.~\"Ovg\"un and D.~Demir,
Class. Quant. Grav. \textbf{40} (2023) No.19, 195003.










\end{thebibliography}
\end{document}